# CONTRASTIVE SELF-SUPERVISED LEARNING FOR SPATIO-TEMPORAL ANALYSIS OF LUNG ULTRASOUND VIDEOS

*Li Chen\*[1], Jonathan Rubin\*†[1], Jiahong Ouyang[1], Naveen Balaraju[1], Shubham Patil[1], Courosh Mehanian[2], Sourabh Kulhare[2], Rachel Millin[2], Kenton W. Gregory[3], Cynthia R. Gregory[3], Meihua Zhu[3], David O. Kessler[4], Laurie Malia[4], Almaz Dessie[4], Joni Rabiner[4], Di Coneybeare[4], Bo Shopsin[5], Andrew Hersh[6], Cristian Madar[7], Jeffrey Shupp[8], Laura S. Johnson[8], Jacob Avila[9], Kristin Dwyer[10], Peter Weimersheimer[11], Balasundar Raju[1], Jochen Kruecker[1], Alvin Chen[1]*

[1] Philips Research North America, [2] Global Health Laboratories, [3] Oregon Health & Science University, [4] Columbia University Medical Center, [5] New York University, [6] Brooke Army Medical Center, [7] Tripler Army Medical Center, [8] MedStar Washington Hospital Center, [9] University of Kentucky, [10] Warren Alpert Medical School of Brown University, [11] University of Vermont Larner College of Medicine

## ABSTRACT

Self-supervised learning (SSL) methods have shown promise for medical imaging applications by learning meaningful visual representations, even when the amount of labeled data is limited. Here, we extend state-of-the-art contrastive learning SSL methods to 2D+time medical ultrasound video data by introducing a modified encoder and augmentation method capable of learning meaningful spatio-temporal representations, without requiring constraints on the input data. We evaluate our method on the challenging clinical task of identifying lung consolidations (an important pathological feature) in ultrasound videos. Using a multi-center dataset of over 27k lung ultrasound videos acquired from over 500 patients, we show that our method can significantly improve performance on downstream localization and classification of lung consolidation. Comparisons against baseline models trained without SSL show that the proposed methods are particularly advantageous when the size of labeled training data is limited (e.g., as little as 5% of the training set).

## 1. INTRODUCTION

Lung ultrasound is an imaging technique deployed to aid in evaluation of lung respiratory diseases, including COVID-19. Pathological features such as consolidations can be visualized under ultrasound [1]–[3], but identifying such features is challenging and requires expertise. Automated identification and visualization of pathology by machine learning models can aid in diagnosis, prognosis, and disease management.

A major challenge in the development of machine learning algorithms on medical images is the need for extensive training data with expert human annotations of high clinical quality. For medical ultrasound in particular, frame-by-frame annotation of a single video can take hours to complete, making annotation-at-scale extremely expensive.

Self-supervised learning (SSL) represents a promising alternative to traditional supervised learning methods. Rather than learning from labels, SSL methods rely on supervisory signals from the underlying unlabeled data to learn latent representations in the absence of ground-truth. State-of-the-art methods based on contrastive learning and other mechanisms of self-supervision [4]–[12] have been reported on 2D natural images. Notably, these methods typically rely on 2D augmentation techniques suited for natural imagery as a means for label-free supervision. Unlike natural images, ultrasound data consists of video sequences with many variables affecting image quality, including scan settings, scan angles, probe motion, body habitus, and other factors.

In this work, we demonstrate that state-of-the-art contrastive learning SSL methods, previously applied only to 2D natural images, may be extended to 2D+time medical ultrasound videos. Specifically, we introduce (1) a modified (3D) encoder to learn spatio-temporal representations, and (2) domain-specific spatio-temporal (3D) augmentations applied in training which simulates the full variability in ultrasound imagery. We show that our adapted SSL method can serve as a pre-training step to initialize weights for downstream localization and video classification tasks, without requiring constraints on the input data. Our method outperforms an equivalent fully-supervised baseline model without SSL pretraining (i.e. initialized with random weights), particularly when the number of labeled samples is extremely limited.

---

## 2. RELATED WORK

The application of SSL for visual tasks has primarily been reported on natural images. Recent works [4]–[6] have shown that well-trained SSL models are competitive with models trained via full supervision. State-of-the-art SSL techniques include contrastive methods utilizing positive-negative pairs (SIMCLR [8] and MoCo [9]); contrastive learning based on asymmetry (BYOL [7] and SIMSIAM [10]); learning visual pretext tasks ([13] and [14]); and learning via redundancy-reduction (Barlow Twins [11] and W-MSE [12]).

SSL methods have been demonstrated on medical images, including ultrasound, for example by leveraging supervision from radiological follow-up scans [15] or reconstructing high-resolution ultrasound images from high- and low-resolution pairs [16]. SSL has also been applied to image synthesis, for example to learn mappings from ultrasound to MR by assuming a shared representation in latent space [17]. More closely related to this work, a self-supervised model was trained to learn visual representations by correcting the order of reshuffled fetal ultrasound videos containing limited numbers of frames per scan and predicting the geometric transformation applied to the videos [18]. Finally, fetal ultrasound imagery was used to train a 2D self-supervised model based on context restoration to facilitate downstream classification, localization, and segmentation [19].

Unlike many of the existing SSL methods applied to medical data, contrastive learning methods, as in [7], [8], do not require specific constraints on the input data, such as needing follow-up scans [15], positive-negative pairs [16], [17], or very short videos (e.g., less than one cardiac cycle) [18]. Instead, contrastive methods leverage asymmetry in the learning update resulting from paired augmentations. For this work, contrastive SSL methods were adapted for 2D+time video through the introduction of domain-specific spatio-temporal augmentations appropriate for medical ultrasound.

## 3. METHODS

### 3.1. Data

An extensive retrospective, multi-center clinical dataset of 27,063 lung ultrasound videos were used in this work (Table 1). The data were acquired from 528 patients with suspicion of lung consolidation or other related pathology (e.g., pneumonia, pleural effusion). The data were collected from 8 U.S. clinical sites between 2017 and 2020. The videos were at least 3 seconds in length and contained at least 60 frames.

To assess model classification performance, 1669 videos were annotated for presence or absence of lung consolidation. Annotation was carried out by a multi-center team of expert physicians with training in lung ultrasound. Each ultrasound video was annotated by two experts and adjudicated by a third expert when disagreement between the first two experts occurred. The annotated videos were then divided at patient level into training, validation, and test sets. The remaining 25,394 videos served as unlabeled data for SSL training.

Table 1 Retrospective, multi-center lung ultrasound dataset.

| Dataset details | N |
| --- | --- |
| Number of sites | 8 |
| Number of patients | 528 |
| Total videos | 27,063 |
| Unlabeled videos | 25,394 |
| Videos labeled with lung consolidation (Train/val/test set) | 1,669 (1,296/120/253) |

### 3.2. Proposed contrastive self-supervised learning method

The proposed contrastive SSL method is shown in Fig. 1. The first step is to generate meaningful visual representations of the ultrasound video so that the visual representations can be used in downstream tasks. In this work, we adopted BYOL ([7]) as a state-of-the-art asymmetry-based contrastive method to learn meaningful visual representations. The success of the method depends on the proper application of two different spatio-temporal augmentation instantiations applied to the same input video during training. The spatio-temporal augmentation parameters are detailed in Table 2.

During training, augmented videos are passed through two neural networks, an online network and a target network. The online network has a 3D encoder (modified Darknet-53 [20] with 3D convolutions as backbone to support video input) to generate visual representations, and a projection head (a multilayer perceptron with one hidden layer of 4096 dimension) to project the embedding features for computation of the loss function. The target network has the same backbone network structure as the online network, but its weights are an exponential moving average (EMA, with momentum update ratio of 0.99) of the online network parameters instead of being backpropagated from later layers.

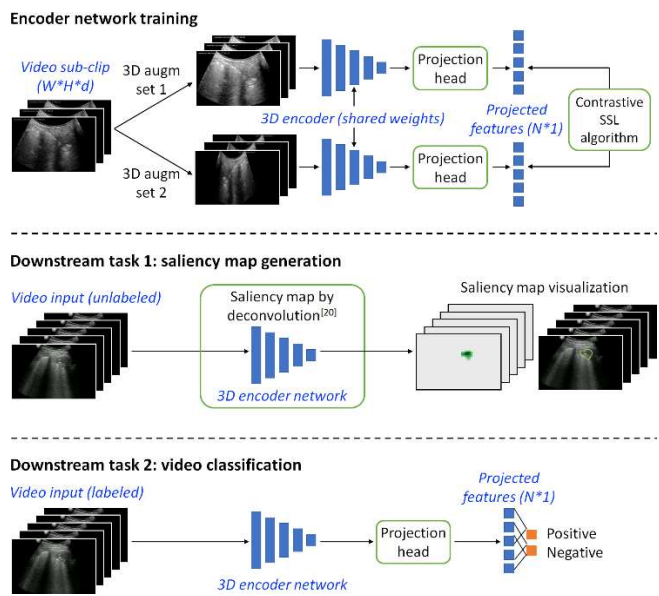

**Fig. 1** Flowchart of proposed contrastive SSL method.

Table 2 Spatio-temporal 2D+time augmentations applied on ultrasound videos in the proposed contrastive SSL method.

| Type | Method | Parameter details |
|---|---|---|
| Spatial augmentations | Random affine transform | Scale ±20%, Translation ±10%, Rotation ±10° |
| | Random horizontal flip | Probability 50% |
| | Random color jitter | Brightness 0.3, Contrast 0.3 |
| | Random Gaussian noise | Standard deviation 0.03 |
| | Random erasing | Proportion (0.02, 0.1) and aspect ratio (0.3, 3.3) of erased areas |
| Temporal augmentations | Reversed frame order | Probability 50% |
| | Shuffled frame order | Probability 50% (0-4 frames) |
| | Random frame replacement | Probability 50% (0-4 frames) |

The parameters of the online network are trained to maximize agreement between the embedding features from both augmented videos. The parameters of the target network are then updated by EMA. After training, the projection head is discarded, and the encoder and its visual representation are used for downstream tasks (e.g., video classification). The learning rate was set to 3e-4 for all experiments.

### 3.3. Self-supervised learning for saliency map generation

As a downstream task, we use the encoder network from SSL training to generate meaningful saliency maps, which tend to highlight regions of each video frame that are likely to contain pathology. In this work, we used one of the popular saliency map generation methods, the Occlusion algorithm [20] on the encoder neural network, although other methods of saliency map generation could be chosen.

To quantify the overlap of saliency map with ground-truth bounding boxes defining regions of pathology (lung consolidation), we use a weighted IOU as the evaluation metric. For this, a threshold mask is applied on the saliency map to retain the top 10% of image pixels based on intensity. Minimum-encompassing prediction boxes are then generated for each connected foreground region and compared with the ground-truth boxes to assess pathology detections.

### 3.4. Self-supervised learning for video classification

We also applied the SSL trained encoder network to the downstream task of video classification. To achieve this, we appended a fully-connected layer (with number of neurons equals to the number of classes) to the visual representation layer within the encoder neural network. The fully-connected layer is subsequently trained on a (smaller) labeled dataset, i.e., via traditional supervised learning.

When training the fully-connected classification layer, the weights of the SSL pre-trained encoder backbone may either remain fixed or allowed to update. We evaluated both approaches in our study. That is, we compared the performance of a classifier in which only the fully-connected layer was tuned based on labeled data ("SSL Feature Extractor") to a classifier in which both the pre-trained encoder network and the fully-connected layer were tuned based on labels ("SSL Fine-Tuned").

We also compared classification performance against an equivalent fully-supervised baseline model without SSL pretraining ("Fully-Supervised"), i.e., initialized with random weights and trained entirely based on labeled data. Finally, to evaluate the importance of the feature extraction layers (encoder network) relative to the fully connected classification layers, we show the results of a naive model with a fixed, random encoder where only the fully-connected layer is trainable ("Random Feature Extractor").

## 4. EXPERIMENTAL RESULTS

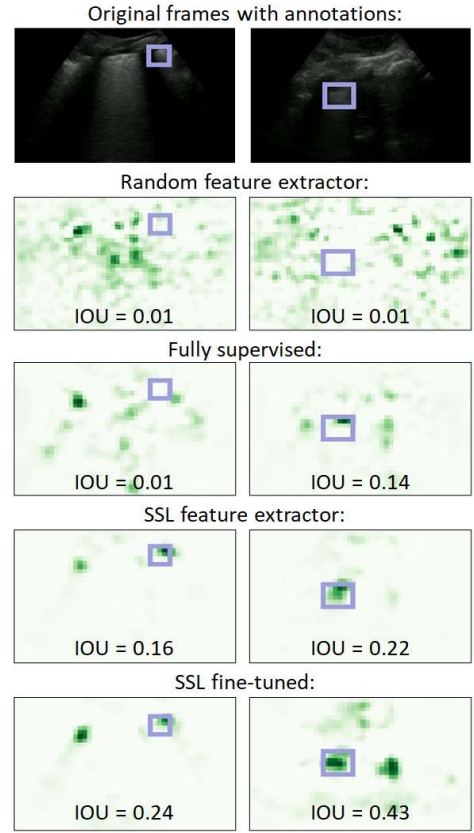

**Fig. 2.** Example frames from lung ultrasound videos with consolidation. Top row shows original frames. Saliency maps (bottom 4 rows) are generated using the method of Zeiler *et al* [20]. Ground truth boxes are shown in purple.

### 4.1. Saliency map generation

Representative examples of SSL generated saliency maps on lung ultrasound videos containing regions of pathology (lung consolidation) are shown in Fig. 2. For these experiments, the SSL models were initially trained using the 25,394 unlabeled videos and 1,269 annotated training videos with fixed ("SSL Feature Extractor") or trainable ("SSL Fine-Tuned") backbone. Baseline models without SSL were trained using only labelled data. As seen in Fig. 2, saliency maps generated by the "Random Feature Extractor" and "Fully-Supervised" models are noisy and cannot clearly localize regions of pathology. In contrast, saliency maps generated from the proposed SSL methods are more specific to the pathology.

### 4.2. Fractional training with limited labeled data

Fig. 3 compares baseline ("Fully-Supervised") and proposed ("SSL Feature Extractor" and "SSL Fine-Tuned") models with decreasing proportional amounts of labeled training data. Specifically, we incrementally reduced the training set from 100% (all 1,296 annotated training videos included) to 5% (65 annotated videos randomly selected for training).

When the amount of labeled training data is sufficient, the baseline "Fully-Supervised" model shows comparable performance to the SSL-based models, and the effect of pre-training with unlabeled data is diminished (0.82 vs 0.84 accuracy, 0.91 vs 0.92 AUC).

On the other hand, when the labeled training set is reduced, the effect of SSL pre-training becomes evident. In particular, we observe that when the proportion of labeled training data falls below 30% of the initial training set size, the accuracy and AUC of the baseline "Fully-Supervised" model decreases dramatically. In contrast, the SSL-based models maintain consistent accuracy and AUC throughout the low data regime, including training with as little as 5% of the annotated dataset.

We also notice that SSL training without temporal augmentations leads to 0.02 decrease in AUC.

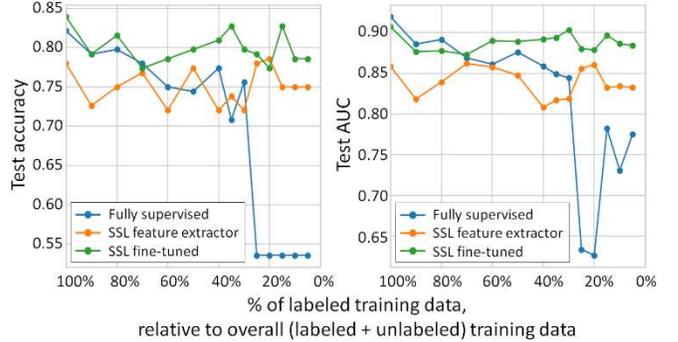

**Fig. 3.** Effect of labeled dataset size on supervised versus self-supervised model performance. When the proportion of labeled training data falls below 30% of the combined labeled and unlabeled training size, the accuracy and AUC of the baseline supervised models decreases dramatically. In contrast, the proposed SSL-based models maintain consistent performance throughout the low data regime.

### 4.3. Comparison with alternative SSL algorithms

To evaluate whether the proposed spatio-temporal augmentation methods may be generalized to other contrastive learning SSL algorithms, we adapt an alternative SSL method [8] (Table 3) with our method. We observe that both algorithms are similar (0.91 vs 0.89 AUC, 0.84 vs 0.79 accuracy) in performance. This suggests that different state-of-the-art contrastive learning techniques may be used with the proposed spatio-temporal augmentation method.

Table 3 Performance comparison between two adapted contrastive learning SSL algorithm applied on lung ultrasound data.

| Model | Trainable parameters (out of 133,880 total) | Accuracy | Sensitivity | Specificity | AUC |
|---|---|---|---|---|---|
| **SSL feature extractor** (Proposed method with SSL algorithm [7]) | 38 | 0.78 | 0.69 | 0.86 | 0.86 |
| **SSL fine-tuned** (Proposed method with SSL algorithm [7]) | 133,880 | **0.84** | **0.74** | 0.92 | **0.91** |
| **SSL feature extractor** (Proposed method with alternative SSL algorithm [8]) | 38 | 0.76 | 0.55 | **0.94** | 0.87 |
| **SSL fine-tuned** (Proposed method with alternative SSL algorithm [8]) | 133,880 | 0.79 | 0.68 | 0.89 | 0.89 |

### 7. CONCLUSIONS

In summary, we extend state-of-the-art contrastive learning SSL methods to 2D+time medical ultrasound video data by introducing a modified encoder and augmentation method to learn meaningful spatio-temporal representations, without added constraints on the input data. We applied the method to the clinically relevant task of video classification of lung consolidations in ultrasound. The results of the study suggest that the proposed SSL methods 1) learn more informative visual representations (saliency maps); 2) outperform baseline models trained without self-supervision; and 3) demonstrate consistent performance even when labeled training data are extremely limited.

## 8. ACKNOWLEDGMENTS

We would like to acknowledge the contributions from the following people for their efforts in data curation and annotations: Zohreh Laverriere, Xinliang Zheng (Lia), Annie Cao, Katelyn Hostetler, Yuan Zhang, Amber Halse, James Jones, Jack Lazar, Devjani Das, Tom Kennedy, Lorraine Ng, Penelope Lema, Nick Avitabile.